\documentstyle[11pt]{article}
\input epsf
\begin{document}
\thispagestyle{empty}
\begin{center}
\null\vspace{-1cm} {\footnotesize
{\tt }}\hfill VACBT/05/13, GNPHE/05/13 and IC/2005\\
\vspace{1cm}
\medskip
\vspace{2.5 cm} {\large \textbf{MOYAL NONCOMMUTATIVE INTEGRABILITY \\AND\\
THE BURGERS-KDV MAPPING}}\\
\vspace{2 cm} \textbf{M.B. SEDRA}\footnote{sedra@ictp.it} \\
{\small \ International Centre for Theoretical Physics,
Trieste, Italy.}\\
{\small \ Virtual African Center For Basic Sciences and
Technology, VACBT,\\ Focal point: Lab/UFR-Physique des Haute
Energies, Facult´e des Sciences,
Rabat, Morocc}.\\
{\small \ Groupement National de Physique de Hautes Energies,
GNPHE, Rabat, Morocco,}\\{\small \ \ \ } {\small \ \ \ }{\small \
Universit\'{e} Ibn Tofail, Facult\'{e} des
Sciences, D\'{e}partement de Physique,}\\
{\small \ \ \ }{\small \ Laboratoire de Physique Th\'{e}orique et
Appliqu\'{e}e (LPTA), K\'{e}nitra, Morocco}\\
\end{center}
\vspace{0.5cm} \centerline{\bf Abstract} \baselineskip=18pt
\bigskip
The Moyal momentum algebra, studied in \cite{ref 20, ref 21}, is
once again used to discuss some important aspects of NC integrable
models and $2d$ conformal field theories. Among the results
presented, we setup algebraic structures and makes useful
convention notations leading to extract non trivial properties of
the Moyal momentum algebra. We study also the Lax pair building
mechanism for particular examples namely, the noncommutative KdV
and Burgers systems. We show in a crucial step that these two
systems are mapped to each others through the following crucial
mapping $\partial_{t_2}\hookrightarrow
\partial_{t_3}\equiv\partial_{t_2}\partial_{x}+\alpha
\partial_{x}^{3}$.
This makes a strong constraint on the NC Burgers system which
corresponds to linearizing its associated differential equation.
From the CFT's point of view, this constraint equation is nothing
but the analogue of the conservation law of the conformal current.
We believe that the considered mapping might help to bring new
insights towards understanding the integrability of noncommutative
$2d$-systems.

\hoffset=-1cm \textwidth=11,5cm \vspace*{1cm}
\hoffset=-1cm\textwidth=11,5cm \vspace*{0.5cm}

\newpage

\section{Introduction}
Recently there has been a revival interest in the noncommutativity
of coordinates in string theory and D-brane physics\cite{ref 1,
ref 2, ref 3, ref 4, ref 5, ref 6}. This interest is known to
concern also noncommutative quantum mechanics and noncommutative
field theories \cite{ref 7, ref 8}. The sharing property between
all the above interesting areas of research is that the
corresponding space exhibits the following structure
\begin{equation}
\begin{array}{lcr}
\lbrack x_{i},x_{j}]_{*^{\prime }}=i\theta _{ij} &  &
\end{array}
\end{equation}
where $x_{i}$ are non-commuting coordinates which can describe
also the space-time coordinates operators and $\theta _{ij}$ is a
constant antisymmetric tensor. Quantum field theories living on
this space are necessarily noncommutative field theories. Their
formulation is simply obtained when the algebra (1) is realized in
the space of fields (functions) by means of the Moyal bracket
according to which the usual product of functions is replaced by
the star-product as follows \cite{ref 9},
\begin{equation}
(f*g)(x) = f(x)e^{\frac{i}{2} \theta^{ab} \overleftarrow{\partial_a}%
\overrightarrow{\partial_b}} g(x) ,
\end{equation}
The link with string theory consist on the correspondence between the $%
\theta^{ij}$-constant parameter and the constant antisymmetric
two-form potential $B^{ij}$ on the brane as follows \cite{ref 1},
\begin{equation}
\theta^{ij} = (\frac{1}{B})^{ij},
\end{equation}
such that in the presence of this $B$-field, the end points of an
open string become noncommutative on the D-brane. The same
interest on the noncommutative geometry is exhibited by the
$(1+1)$-dimensional integrable models \cite{ref 10} which are
intimately connected to conformal field theories \cite{ref 11} and
their underlying lower $(s\leq 2)$ \cite{ref 12,ref 13,ref 14,ref
15, ref 16} and higher $(s\geq 2)$ \cite{ref 17, ref 18} spin
symmetries.\\\\ Non trivial integrable models are, in general
situations, based on nonlinear differential equations which may be
solvable. However, few nonlinear differential equations which are
integrable. This property is traced to the fact that solving a
nonlinear system is not an easy job in most of the considered
cases.\\\\
As the notion of integrability, of a given system, is one of
important physical requirements, one have to overcome the
difficulties of nonlinearity by adopting adequate technics. In
this sense, one can anticipate a first definition of integrability
as been the possibility to linearize the associated nonlinear
differential equation. The famous approach of linearizing a
physical system is given by the well known Lax technics related to
the inverse scattering transformation\cite{ref 19}. The idea
consist in assuming the existence of a pair of operators, $L$ and
$B$ satisfying the following linear equation
\begin{equation}
{\partial_{t} \mathcal L}=[{\mathcal L}, B].
\end{equation}
Usually the application of the inverse scattering transformation
method to an evolution equation is based on the Lax
representation. The linear Lax equation (4) describes then an
evolution equation of the differential operators $L$ and $B$ with
$[\mathcal L ;B]$ is their commutator.\\\\ One of the principal
objectives of this work is study the Lax representation in the
case of noncommutative systems. This is an important mathematical
and physical issue expected to shed more lights on the notion of
integrability. We will make some consistent assumptions shown to
be essential in in deriving the Lax pair of special noncommutative
integrable systems.\\\\ One need to extract new properties of the
integrability in the case of noncommutative spaces. The Lax
representation provides a sophisticated means to achieve such a
goal. It's commonly known that the existence of the Lax pair gives
a significant sign of integrability. The importance of
noncommutative extensions find then a good motivation through this
issue. Accordingly, we will consider Lax formalism as the way to
test the integrability of the noncommutative systems. In the same
philosophy, the intervention of conformal symmetry is planned to
reinforce our analysis and to push our research of the source of
the integrability ahead.\\\\The principal prototype examples
considered in this study are the KdV and Burgers systems. The
originality of the present work deals with the possibility to
connect these noncommutative systems through a consistent mapping
that we will setup. Before describing the essential of this
mapping transformation, let's first present briefly the content
the successive sections.\\\\
We give in \emph{section{2}} some generalities on the convention
notations that we use and on the algebraic structures of the Moyal
momentum algebra ${\widehat \Sigma}$ introduced in  \cite{ref 20,
ref 21}.\\\\ In \emph{section {3}} we present an explicit
description of the Moyal momentum algebra ${\widehat \Sigma}$ and
show how it can relies to $2d$-CFT. \emph{section{4}} is devoted
to some important implications of the Moyal momentum algebra on
generalized KdV hierarchies. We will concentrate on the
noncommutative $sl_2$ KdV equation and the noncommutative version
of the Burgers equation and their Lax representations. The derived
properties may naturally be derived for the more general case
namely the $sl_n$ KdV hierarchy.\\\\
The \emph{Section 5} is devoted to a set up of the Lax pair
representation of special noncommutative integrable systems namely
the noncommutative KdV and Burgers systems. Here we present a
systematic study of noncommutative generating Lax pair operators
in the Moyal momentum framework. The essential results deals with
the derivation of the noncommutative KdV and Burgers systems.
Concerning the noncommutative derived KdV system, this is an
integrable model due to its underlying conformal symmetry.
\\\\ Going in the same lines of our previous works on noncommutative
geometry $\grave a$ la Moyal, we try in \emph{section 6} to study
a possible relation between the NC KdV and the NC Burgers systems.
Several important facts to support this possibility are
discussed.\\
\section{Noncommutativity $\grave a$ la Moyal: Generalities}
\subsection {Basic Definitions}

{\bf 1}. We  start first by recalling that the functions often
involved in the $2d$-phase-space are arbitrary functions which we
generally indicate by $f(x,p)$ where the variable $x$ stands for
the space coordinate
while $p$ describes the momentum coordinate. \\
{\bf 2}. With respect to this phase space, we have to precise that
the constants $f_0$ are defined such that
\begin{equation}
 \partial_x f_0=0= \partial_p {f_0}.
\end{equation}\\
{\bf 3}. The functions ${u_i}(x,t)$ depending on an infinite set
of variables $t_1 =x, t_2,t_3,...,$ do not depend on momentum
coordinates, which means
\begin{equation}
\partial_{p} u_{i} (x,t)=0,
\end{equation}
where the index $i$, describes the conformal weight of the field
$u_i (x,t)$. These functions can be considered in the complex
language framework as
being the analytic (conformal) fields of conformal spin $i=1,2,...$. \\
{\bf 4}. Other objects usually used are the ones given by
\begin{equation}
u_i (x,t)\star p^j,
\end{equation}
which are objects of conformal weight $(i+j)$ living on the
non-commutative space parametrized by $\theta$. Through this work,
we will use the following convention notations $[u_i]= i$,
$[\theta] =0$ and  $[p]=[\partial _x]=-[x]=1$,
where the symbol $[\hspace{0,5 cm}]$ stands for the conformal dimension of the used objects.\\
{\bf 5}. The star product law, defining the multiplication of
objects in the non-commutative phase space, is given by the
following expression
\begin{equation}
f(x,p)\star g(x,p)= \sum_{s=0}^{\infty}
\sum_{i=0}^{s}{\frac{\theta ^s}{s!}} (-)^{i} c _{s}^{i}
(\partial_{x}^{i}\partial_{p}^{s-i}f)(\partial_{x}^{s-i}\partial_{p}^{i}g),
\end{equation}
with $c _{s}^{i}=\frac {s!}{i!(s-i)!} $.\\
{\bf 6}. The conventional Moyal bracket is defined as
\begin{equation}
\{f(x,p), g(x,p)\}_ {\theta} =\frac {f \star g - g \star
f}{2\theta},
\end{equation}
where $\theta$ is the noncommutative parameter, considered as a
constant in this approach{\footnote{For an application of non
constant $\theta$ parameter, see for instance \cite{ref 22} and
references therein}}.\\
{\bf 7}. To distinguish the classical objects from the
$\theta$-deformed ones, we consider the following convention notations \cite{ref 20, ref 21}:\\
a)$\widehat{\Sigma} _{m}^{(r,s)}$: This is the space of momentum
differential operators of conformal weight $m$ and degrees $(r,s)$
with $r\leq s$. Typical operators of this space are given by
\begin{equation}
\sum _{i=r}^{s}u_{m-i}\star p^{i}.
\end{equation}\\
b)$\widehat{\Sigma} _{m}^{(0,0)}$: This is the space of functions
of conformal weight $m$; $ m\in Z$, which may depend on the
parameter $\theta$. It coincides in the classical limit, $\theta
\rightarrow \theta_{l}$\footnote{Usually the standard limit is
taken such that $\theta_{limit}=0$. In the present analysis, the
standard limit is shifted by $\frac{1}{2}$ such that
$\theta_{l}\rightarrow\theta_{limit}+\frac{1}{2}$. Thus taking the
standard limit to be $\theta_{limit}=0$ is equivalent to set
$\theta_{l}=\frac{1}{2}$. The origin of this shift belongs to the
consistent non commutative $w_{\theta}^{3}$-Zamolodchikov algebra
construction \cite{ref 20, ref 21}}\\, with the ring of analytic
fields involved into the construction
of conformal symmetry and $w$-extensions.\\
c)${\widehat\Sigma}_{m}^{(k,k)}$: Is the space of momentum
operators type,
\begin{equation}
u_{m-k}\star p^{k}.
\end{equation}\\
d)$\theta$-\textbf{Residue operation:} $\widehat{Res}$
\begin{equation}
\widehat{Res}(f\star p^{-1})=f.
\end{equation}
\subsection {The Moyal Momentum algebra ${\widehat
\Sigma} (\theta)$:} This is the algebra based on arbitrary
momentum differential operators of arbitrary conformal weight $m$
and arbitrary degrees $(r,s)$. Its obtained by summing over all
the allowed values of spin (conformal weight) and degrees in the
following way:
\begin{equation}
{\widehat \Sigma} (\theta) = \oplus _{r\leq s} \oplus _{m \in
Z}{\widehat \Sigma}_{m}^{(r,s)}.
\end{equation}
${\widehat \Sigma} (\theta)$ is an infinite dimensional momentum
algebra which is closed under the Moyal bracket without any
condition. A remarkable property of this space is the possibility
to introduce six infinite dimensional classes of momentum
sub-algebras related to each other by special duality relations.
These classes of algebras are given by ${\widehat
\Sigma}_{s}^{\pm},$ with $s=0,+,-$ describing respectively the
different values of the conformal spin which can be zero, positive
or negative. The $\pm$ upper indices stand for the values of the
degrees quantum numbers, for more details see \cite{ref 16, ref 20, ref 21}.\\

\subsubsection{Algebraic structure of The space ${\widehat \Sigma}
_{m}^{(r,s)}$} To start let's precise that this space contains
momentum operators of fixed conformal spin m and degrees (r,s),
type
\begin{equation}
{\mathcal {L}}_{m}^{(r,s)}(u)=\sum _{i=r}^{s}u_{m-i}(x)\star p^i,
\end{equation}
These are $\theta$-differentials whose operator character is
inherited from the star product law defined as in (8).\\
Using this relation, it is now important to precise how the
momentum
operators act on arbitrary functions $f(x,p)$ via the star product.\\
Performing computations based on relation (8), we find the
following $\theta$- Leibnitz rules:
\begin{equation}
p^{n} \star f(x,p) = \sum _{s=0}^{n} \theta ^{s} c_{n}^{s}
f^{(s)}(x,p) p^{n-s},
\end{equation}
and
\begin{equation}
p^{-n} \star f(x,p) = \sum _{s=0}^{\infty} (-)^{s} \theta ^{s}
c_{n+s-1}^{s} f^{(s)}(x,p) p^{-n-s},
\end{equation}
where $f^{(s)}=\partial_{x}^{s}f$ is the prime derivative. We also
find the following expressions for the Moyal bracket:
\begin{equation}
\begin{array}{lcl}
\{p^n, f\}_{\theta} &=&\sum _{s=0}^{n} \theta ^{s-1} c_{n}^{s}\{\frac {1-(-)^s}{2}\} f^{s} p^{n-s},\\\\
\{p^{-n}, f\}_{\theta} &=& \sum _{s=0}^{\infty} \theta ^{s-1}
c_{s+n-1}^{s}\{\frac {(-)^{s}-1}{2}\} f^{s} p^{-n-s},
\end{array}
\end{equation}
Special Moyal brackets are given by
\begin{equation}
\begin{array}{lcl}
\{p, x\}_{\theta} &=&1\\\\
\{p^{-1}, x\}_{\theta} &=&-p^{-2}
\end{array}
\end{equation}
With the derived the Leibnitz rules for the momentum operators, we
can also remark that the momentum operators $p^i$ satisfy the
algebra
\begin{equation}
p^n \star p^m =p^{n+m}.
\end{equation}
which ensures the suspected rule
\begin{equation}
\begin{array}{lcl}
p^{n} \star (p^{-n}\star f) &=& f\\\\
(f\star p^{-n} )\star p^{n}&=&f.
\end{array}
\end{equation}
\subsubsection{Further Algebraic Properties of ${\widehat
\Sigma}_{m}^{(r,s)}$:}

An important algebraic property of the space ${\widehat
\Sigma}_{m}^{(r,s)}$ is that it may decomposes into the underlying
subspaces as
\begin{equation}
{\widehat \Sigma}_{m}^{(r,s)} =\oplus _{k=r}^{s} {\widehat \Sigma}%
_{m}^{(k,k)}(\theta) \\
\end{equation}
where ${\widehat \Sigma}_{m}^{(k,k)}$ are unidimensional subspaces
containing prototype elements of kind $u_{m-k}\star p^{k}$ or
$p^{k}\star u_{m-k}$. Using the $\theta$-Leibniz rule, we can
write, for fixed value of k:
\begin{equation}
{\widehat \Sigma}_{m}^{(k,k)} \equiv \Sigma _{m}^{(k,k)} \oplus
\theta \Sigma _{m}^{(k-1,k-1)} \oplus
\theta^{2}{\Sigma}_{m}^{(k-2,k-2)} \oplus...
\end{equation}
where $\Sigma _{m}^{(k,k)}$ is the standard one dimensional
sub-space of Laurent series objects $u_{m-k} p^k$ considered also as the $(\theta=0)$%
-limit of ${\widehat \Sigma}_{m}^{(k,k)}$. \newline
\newline
This property can be summarized as follows
\begin{equation}
\begin{array}{lcl}
{\widehat \Sigma}_{m}^{(r,s)} & = & \oplus _{k=r}^{s} {\widehat \Sigma}%
_{m}^{(k,k)}(\theta) \\
&  &  \\
& = & \oplus _{k=r}^{s}\oplus _{l=0}^{k}
{\theta}^{l}{\Sigma}_{m}^{(k-l,k-l)}
\end{array}
\end{equation}
Furthermore, the unidimensional subspaces ${\widehat
\Sigma}_{m}^{(k,k)}$ can be written formally as
\begin{equation}
{\widehat \Sigma}_{m}^{(k,k)} \equiv p^{k} \star \Sigma
_{m-k}^{(0,0)}.
\end{equation}
where ${\widehat \Sigma}_{m}^{(0,0)} \equiv \Sigma _{m}^{(0,0)}$
is nothing but the ring of analytic fields $u_{m}$ of conformal
spin ${m \in Z}$ satisfying
\begin{equation}
u_i\star u_j=u_{i}. u_{j}
\end{equation}
Another property concerning the space ${\widehat
\Sigma}_{m}^{(r,s)}$ is its non closure under the action of the
Moyal bracket since we have;
\begin{equation}
\{.,.\}_\theta : {\widehat \Sigma}_{m}^{(r,s)} \star {\widehat \Sigma}%
_{m}^{(r,s)} \rightarrow{\widehat \Sigma}_{2m}^{(r,2s-1)}
\end{equation}
Imposing the closure, one gets strong constraints on the integers
$m$, $r$ and $s$ namely
\begin{equation}
\begin{array}{lcl}
m & = & 0 \\
r \leq & s & \leq 1
\end{array}
\end{equation}
With these constraint equations, the sub-spaces ${\widehat \Sigma}%
_{m}^{(r,s)}$ exhibit then a Lie algebra structure since the
$\star$-product is associative.\\\\

\subsubsection{Residue Duality in ${\widehat \Sigma}_{m}^{(r,s)}$:}

The sub-space ${\widehat \Sigma}_{m}^{(r,s)}$ is characterized by
the existence of a residue operation that we denote as ${\widehat
Res}$ and which acts as follows
\begin{equation}
\begin{array}{lcl}
\widehat{Res}(u_{k}\star p^{-k}) & = & (u_{k} \star p^{-k}) \delta
_{k-1,0}
\\
&  &  \\
& = & u_{1}\delta _{k-1,0}
\end{array}
\end{equation}
This result coincides with the standard residue operation: $Res$,
acting on the sub-space ${\Sigma}_{m}^{(r,s)}$:
\begin{equation}
Res (u_1 . p^{-1})=u_1
\end{equation}
We thus have two type of residues $\widehat Res$ and $Res$ acting
on two different spaces ${\widehat \Sigma}_{m}^{(r,s)}$ and
${\Sigma}_{m}^{(r,s)}$ but with value on the same ring
${\Sigma}_{m+1}^{(0,0)}$. This Property is summarized as follows:
\begin{equation}
\begin{array}{lcl}
{\widehat \Sigma}_{m}^{(r,s)} \hspace{0,5cm}\stackrel{\small \theta =0}{%
\longrightarrow} \hspace{0,5cm} {\Sigma _{m}^{(r,s)}} &  &  \\
\hspace{0,2cm}{_{{\small \widehat Res}}} {\large \searrow}
\hspace{1cm}
{\large \swarrow}_{{\small {Res}}} &  &  \\
\hspace{1,15cm} \Sigma _{m+1}^{(0,0)} &  &
\end{array}
\end{equation}
We learn from this diagram that the residue operation exhibits a
conformal spin quantum number equal to 1 as it maps objects of
conformal spin $m$ to the space ${\widehat \Sigma}_{m+1}^{(0,0)}$.
The other important property of the residue operation is that it
acts only on the ${\widehat \Sigma}_{m}^{(-1,-1)}$\newline
With respect to the previous residue operation, we define on ${\widehat{%
\Sigma }}$ the following degrees pairing product
\begin{equation}
\left( .,.\right) :{\widehat{\Sigma }}_{m}^{(r,s)}\star {\widehat{\Sigma }}%
_{n}^{(-s-1,-r-1)}\rightarrow {\Sigma }_{m+n+1}^{(0,0)}
\end{equation}
such that
\begin{equation}
\left( {\cal L}{}_{m}^{(r,s)}(u),\tilde{{\cal L}}_{n}^{(\alpha
,\beta
)}(v)\right) =\delta _{\alpha +s+1,0}\delta _{\beta +r+1,0}{\widehat{R}es}%
\left[ {\cal L}{}_{m}^{(r,s)}(u)\star \tilde{{\cal
L}}_{n}^{(\alpha ,\beta )}(v)\right] ,
\end{equation}
showing that the spaces ${\widehat{\Sigma }}_{m}^{(r,s)}$ and ${\widehat{%
\Sigma }}_{n}^{(-s-1,-r-1)}$ are $\widehat{R}es$-dual as ${\Sigma }%
_{m}^{(r,s)}$ and ${\Sigma }_{n}^{(-s-1,-r-1)}$ are dual with
respect to the standard $Res$-operation.\newline

\subsubsection{The Lie algebra ${\widehat\Sigma}_{0}^{(-\infty,1)}$
section of ${\widehat{\Sigma}}_{m}^{(r,s)}$} As discussed
previously, the subspaces ${\widehat{\Sigma}}_{m}^{(r,s)}$ exhibit
a Lie algebra structure with respect to the Moyal bracket once the
spin-degrees constraints (27) are considered. With these
conditions one should note that the huge Lie algebra that we can
extract from the space ${\widehat{\Sigma }}_{m}^{(r,s)}$ consists
on the space ${\widehat{\Sigma }}_{0}^{(-\infty ,1)}$ having the
remarkable space decomposition
\begin{equation}
{\widehat{\Sigma }}_{0}^{(-\infty ,1)}={\widehat{\Sigma
}}_{0}^{(-\infty ,-1)}\oplus {\widehat{\Sigma }}_{0}^{(0,1)},
\end{equation}
where ${\widehat{\Sigma }}_{0}^{(-\infty ,-1)}$ describes the Lie
algebra of pure non local momentum operators and ${\widehat{\Sigma
}}_{0}^{(0,1)}$ is
the Lie algebra of local Lorentz scalar momentum operators ${\cal L}$ $%
{}_{0}(u)=u_{-1}\star p+u_{0}$. The latter can splits as follows
\begin{equation}
{\widehat{\Sigma }}_{0}^{(0,1)}={\widehat{\Sigma }}_{0}^{(0,0)}\oplus {%
\widehat{\Sigma }}_{0}^{(1,1)},
\end{equation}
where ${\widehat{\Sigma }}_{0}^{(1,1)}$ is the Lie algebra of
vector momentum fields $J_{0}(u)=u_{-1}\star p$ which are also
elements of ${\Sigma }_{0}^{(0,1)}$.\newline
\newline
As a prototype example, consider
\begin{equation}
\begin{array}{lcl}
{}{\cal L}_{u} & = & u_{-1}\star p+u_{0} \\
&  &  \\
{\cal L}{}_{v} & = & v_{-1}\star p+v_{0}
\end{array}
\end{equation}
be two elements of ${\widehat{\Sigma }}_{0}^{(0,1)}$.
Straightforward computations lead to:
\begin{equation}
\{{\cal L}_{u}, {{\cal L}_{v}}\}_{\theta }={\cal L}_{w}
\end{equation}
with
\begin{equation}
\begin{array}{lcl}
{}{\cal L}_{w} & = & w_{-1}\star p+w_{0} \\
&  &  \\
& = & \{u_{-1}v_{-1}^{\prime }-u_{-1}^{\prime }v_{-1}\}\star
p+\{u_{-1}v_{0}^{\prime }-u_{0}^{\prime }v_{-1}\}.
\end{array}
\end{equation}
Forgetting about the fields ({\it of vanishing conformal spin})
belonging to ${\Sigma }_{0}^{(0,0)}$ is equivalent to consider the
coset space
\begin{equation}
{\widehat{\Sigma }}_{0}^{(1,1)}\equiv {{\widehat{\Sigma }}_{0}^{(0,1)}}/{{%
\Sigma }_{0}^{(0,0)}}
\end{equation}
one obtain the $Diff(S^{1})$ momentum algebra of vector fields $%
J_{0}(u)=u_{-1}\star p$ namely
\begin{equation}
\{J_{0}(u),J_{0}(v)\}_{\theta }=J_{0}(w)
\end{equation}
with $w_{-1}=u_{-1}v_{-1}^{\prime }-u_{-1}^{\prime
}v_{-1}$.\newline

The extension of these results to non local momentum operators is
natural. In fact, one easily show that the previous Lie algebras
are simply
sub-algebras of the huge momentum space ${\widehat \Sigma}_{0}^{(-\infty,1)}$%
. For a given $0\leq k\leq 1$, we have
\begin{equation}
{\widehat \Sigma}_{0}^{(0, 1)}\subset {\widehat
\Sigma}_{0}^{(-\infty, k)}\subset {\widehat \Sigma}_{0}^{(-\infty,
1)}
\end{equation}
and by virtue of (47)
\begin{equation}
\{{\widehat \Sigma}_{0}^{(-\infty,k)}, {\widehat \Sigma}_{0}^{(0,
1)}\}_{\theta}\subset {\widehat \Sigma}_{0}^{(-\infty, k)}\subset
{\widehat \Sigma}_{0}^{(-\infty,1)}
\end{equation}
and for $-\infty < p \leq q\leq 1$
\begin{equation}
\{{\widehat \Sigma}_{0}^{(-\infty, p)}, {\widehat
\Sigma}_{0}^{(-\infty, q)}\}_{\theta}\subset {\widehat
\Sigma}_{0}^{(-\infty, p+q-1)}
\end{equation}
These Moyal bracket expressions show in turn that all the subspaces ${%
\widehat \Sigma}_{0}^{(p,q)}$ with $-\infty < p \leq q\leq 1$ are
ideals of of ${\widehat \Sigma}_{0}^{(-\infty,1)}$.
\section{The $sl_n-\widehat \Sigma_{n}^{(0,n)}(\theta)$ NC Algebra and
$2d$ CFT}

The algebra $sl_n-\widehat \Sigma_{n}^{(0,n)}(\theta)$ describes
simply the coset space
${\widehat{\Sigma }}_{n}^{(0,n)}/{\widehat{%
\Sigma }}_{n}^{(1,1)}$ of $sl_{n}$-Lax operators given by
\begin{equation}
{{}{\cal L}_{n}}(u)=p^{n}+\sum_{i=0}^{n-2}u_{n-i}\star p^{i}
\end{equation}
where we have set $u_{0}=1$ and $u_{1}=0$. This is a natural
generalization of the well known $sl_{2}$-momentum Lax operator
\begin{equation}
{}{\cal L}_{2}={p}^{2}+u_{2}
\end{equation}
associated to the $\theta $-KdV integrable hierarchy that we will
discuss later.\newline
\newline
We have to underline that the $sl_{n}$-momentum Lax operators play
a central role in the study of integrable models and more
particularly in deriving higher conformal spin algebras
($w_{\theta }$-algebras) from the $\theta $-extended
Gelfand-Dickey second Hamiltonian structure \cite{ref 20, ref 21,
ref 23}. Since they are also important in recovering $2d$
conformal field theories via the Miura transformation, we guess
that its possible to extend this property, in a natural way, to
the non-commutative case and consider the $\theta $-deformed
analogue of the well known $2d$ conformal models namely: the
$sl_{2}$-Liouville field theory and its $sl_{n}$-Toda extensions
and also the Wess-Zumino- Novikov-Witten conformal model.\newline
\newline
As an example consider the $\theta $-KdV momentum Lax operator
that we can write as
\begin{equation}
\begin{array}{lcl}
{\cal L}{}_{2} & = & {p}^{2}+u_{2} \\
& = & (p+{\phi }^{\prime })\star (p-{\phi }^{\prime })
\end{array}
\end{equation}
where $\phi $ is a Lorentz scalar field. As a result we have
\begin{equation}
u_{2}=-{\phi ^{\prime }}^{2}-2{\theta }{\phi }^{\prime \prime }
\end{equation}
which is nothing but the $\theta $ analogue of classical stress
energy momentum tensor of $2d$ conformal Liouville field theory.
Using $2d$ complex coordinates language, we can write
\begin{equation}
T_{\theta }(z)\equiv u_{2}(z)=-2\theta {\partial ^{2}}\phi
-(\partial \phi )^{2}
\end{equation}
with $\partial \equiv \partial _{z}\equiv \frac{\partial
}{\partial z}$. The conservation for this conformal current namely
$\bar{\partial}T(z)=0$, leads to write the following $\theta
$-Liouville equation of motion
\begin{equation}
\partial {\bar{\partial}}\phi =\frac{2}{\theta }e^{-\frac{1}{\theta }\phi }
\end{equation}
associated to the two dimensional $\theta $-Liouville action
\begin{equation}
S=\int {d^{2}z\left( \frac{1}{2}{\partial \phi }\star {\bar{\partial}\phi }%
+e^{-\frac{1}{\theta }\phi }\right) }
\end{equation}
with ${\partial \phi }\star {\bar{\partial}\phi }={\partial \phi }{\bar{%
\partial}\phi }$.\newline
\newline
Note by the way that we may interpret the inverse $\frac{1}{\theta
}$ of the non-commutative $\theta $-parameter as being the
analogue of the Cartan matrix of $sl_{2}$ because in the classical
limit this Cartan matrix is known to be $(a_{ij})=2$ once the
classical limit $\theta_{l}\rightarrow \frac{1}{2}$ is
considered.\newline Another important point is that we know from
the standard $2d$ CFT \cite{ref 11} that the object $T_{zz}$
satisfying
\begin{equation}
T_{zz}=-\frac{1}{4}(\partial \phi )^{2}+i\alpha _{0}{\partial
^{2}}\phi
\end{equation}
is nothing but the Feigin-Fuchs representation of the conserved
current generating the conformal invariance of a quantum conformal
model with the central charge $c=(1-24{\alpha _{0}}^{2})$.\newline
\newline
Using this standard result, we can conclude that the $\theta
$-conformal current that we derive in (47) (with the rescaling $\phi \equiv \frac{1}{2}%
\phi $) is associated to the $\theta $-Liouville model having the
central charge
\begin{equation}
c_{\theta }=(1+24{\theta }^{2})
\end{equation}
where the non-commutative parameter is shown to coincide with
$\alpha _{0}$ as follows $\theta =-i\alpha _{0}$.\newline
\newline
The analysis that we use to derive the $\theta $-Liouville
equation and its central charge a la Feigin-Fuchs, can be
generalized to higher conformal
spin Toda field theories associated to $sl_{n}$ symmetry with $(n-1)-$%
conserved currents $T(z),w_{3},w_{4},...w_{n}$. \newline
\newline
Finally we note that all the properties discussed above may be
generalized to the $sl_{n}$ case. This is an explicit proof of the
importance of the algebraic structure inherited from the Moyal
momentum algebra. Actually we showed how these algebra may leads
to extend, in a successful, way all the important properties of
$2d$ CFT theories. We will present in
the next section some other applications of the momentum algebra in $\theta $%
-integrable KdV hierarchies
\section{Noncommutative $sl_n$ KdV-hierarchy}

The aim of this section is to present some results related to the $\theta$%
-KdV hierarchy and that are explicitly derived in \cite{ref 20,
ref 21}. Using our convention notations and the analysis that we
developed previously, we will perform hard algebraic computations
and derive the $\theta$-KdV hierarchy.
\newline
The concerned computations are very hard and difficult to realize
in the general case. For this reason, we concentrate simply on the
first orders of the hierarchy namely the $sl_2$-KdV and $sl_3$-Boussinesq $\theta$%
-integrable hierarchies.\newline
\newline
The study concerns some results obtained in \cite{ref 20, ref 21}
generalizing the ones obtained in \cite{ref 23} by increasing the
order of computations a fact which leads to discover more
important properties. As an original result, we were able to build
the $\theta$-deformed $sl_3$-Boussinesq hierarchy and derive the
associated $\theta$-flows.

\subsection{$sl_2$-KdV hierarchy}

Let's consider the $sl_{2}$-momentum Lax operator
\begin{equation}
{\cal L}{}_{2}=p^{2}+u_{2}
\end{equation}
whose 2th root is given by
\begin{equation}
\begin{array}{lcl}
{}{\cal L}^{\frac{1}{2}} & = & \Sigma _{i=-1}b_{i+1}\star p^{-i} \\
&  &  \\
& = & {\Sigma }_{i=-1}a_{i+1}p^{-i}
\end{array}
\end{equation}
This 2th root of ${\cal L}{}_{2}$ is an object of conformal spin $[{\cal L}%
{}^{\frac{1}{2}}]=1$ that plays a central role in the derivation of the $%
\theta $-Lax evolutions equations. In the spirit to contribute
much more to this $sl_{2}$-KdV hierarchy, it was important for us
to make contact with previous works in literature \cite{ref 23}.
\newline Performing lengthy but straightforward calculations
we compute the coefficients $b_{i+1}$ of ${\cal
L}{}^{\frac{1}{2}}$ up to $i=7$ given by
\footnote{%
The authors of \cite{ref
23} present explicit computations of the coefficients $%
a_{i+1}$ and omit the $b_{i+1}$ ones. Here, we give explicit
computation of both of them}:
\begin{equation}
\begin{array}{lcl}
b_{0} & = & 1 \\
&  &  \\
b_{1} & = & 0 \\
&  &  \\
b_{2} & = & \frac{1}{2}u \\
&  &  \\
b_{3} & = & -\frac{1}{2}\theta u^{^{\prime }} \\
&  &  \\
b_{4} & = & -\frac{1}{8}u^{2}+\frac{1}{2}\theta ^{2}u^{^{\prime \prime }} \\
&  &  \\
b_{5} & = & -\frac{1}{2}\theta ^{3}u^{^{\prime \prime \prime }}+\frac{3}{4}%
\theta uu^{^{\prime }} \\
&  &  \\
b_{6} & = & \frac{1}{16}u^{3}-\frac{7}{4}\theta ^{2}uu^{^{\prime \prime }}-%
\frac{11}{8}{\theta }^{2}(u^{^{\prime }})^{2}+\frac{1}{2}{\theta }%
^{4}u^{^{\prime \prime \prime \prime }} \\
&  &  \\
b_{7} & = & -\frac{15}{16}\theta u^{2}u^{\prime }+{\theta }^{3}(\frac{15}{2}%
u^{\prime \prime }u^{\prime }+\frac{15}{4}uu^{\prime \prime \prime \prime })-%
\frac{1}{2}{\theta }^{5}u^{(5)} \\
&  &  \\
b_{8} & = & -\frac{5}{128}u^{4}+{\theta
}^{2}(\frac{55}{16}u^{\prime \prime
}u^{2}+\frac{85}{16}u{u^{\prime }}^{2})-{\theta
}^{4}(\frac{31}{4}uu^{\prime
\prime \prime \prime }+\frac{91}{8}u^{\prime \prime }{}^{2}+\frac{37}{2}%
u^{\prime }u^{\prime \prime \prime })+\frac{1}{2}{\theta
}^{6}u^{(6)}
\end{array}
\end{equation}
and
\begin{equation}
\begin{array}{lcl}
b_{9} & = & \frac{35}{32}\theta u^{3}u^{\prime }-\frac{175}{4}{\theta }%
^{3}\left( uu^{\prime }u^{\prime \prime }+\frac{1}{4}u^{\prime }{}^{3}+\frac{%
1}{4}u^{2}u^{\prime \prime \prime }\right) +\frac{7}{4}{\theta
}^{5}\left( 9uu^{(5)}+25u^{(4)}u^{\prime }+35u^{\prime \prime
\prime }u^{\prime \prime
}\right) \\
&  &  \\
&  & -\frac{1}{2}\theta ^{7}u^{(7)} \\
&  &  \\
b_{10} & = & {\frac{7}{256}}u^{5}-\frac{35}{32}{\theta }^{2}\left( \frac{23}{%
2}u^{2}u^{^{\prime }2}+5u^{3}u^{\prime \prime }\right) +\frac{7}{4}{\theta }%
^{4}\left( \frac{73}{4}u^{2}u^{(4)}+\frac{227}{4}uu^{^{\prime \prime }2}+%
\frac{337}{4}u^{\prime \prime }u^{^{\prime }2}+89uu^{\prime
}u^{\prime
\prime \prime }\right) \\
&  &  \\
&  & -\frac{3}{4}{\theta }^{6}\left( \frac{631}{3}u^{^{\prime
\prime
}}u^{(4)}+233u{u^{\prime \prime \prime }}^{2}+135u^{\prime }u^{(5)}\right) +%
\frac{1}{2}\theta ^{8}u^{(8)} \\
&  &
\end{array}
\end{equation}
These results are obtained by using the identification ${\cal L}{}_{2}={\cal %
L}{}^{\frac{1}{2}}*{\cal L}{}^{\frac{1}{2}}$. Note that by virtue
of (53), the coefficients $a_{i+1}$ are shown to be functions of
$b_{i+1}$ and their derivatives in the following way
\begin{equation}
a_{i+1}=\sum_{s=0}^{i-1}{\theta ^{s}}c_{i-1}^{s}b_{i+1-s}^{(s)},
\end{equation}
Substituting the derived expressions of $b_{i+1}$ (54-55) into
(56), we obtain the results presented in \cite{ref 23} namely:\footnote{%
In this work, important explicit computations of the parameters
$a_{i+1}$ are presented up to $a_{10}$. Our calculus
(57-58)\cite{ref 20, ref 21}, performed up to $a_{12}$ show some
missing terms in the computations of \cite{ref 23} relative to
$a_{10}$}
\begin{equation}
\begin{array}{lcl}
a_{0} & = & 1 \\
&  &  \\
a_{2} & = & \frac{1}{2}u \\
&  &  \\
a_{4} & = & -\frac{1}{8}u^{2} \\
&  &  \\
a_{6} & = & \frac{1}{16}u^{3}+\frac{1}{8}{\theta }^{2}(u^{\prime
}{}^{2}-2uu^{^{\prime \prime }}) \\
&  &  \\
a_{8} & = & -{\frac{5}{128}}u^{4}+{\frac{5}{8}}{\theta }^{2}\left(
u^{2}u^{\prime \prime }-\frac{1}{2}u^{^{\prime }2}u\right) +{\frac{1}{4}}{%
\theta }^{4}\left( u^{\prime \prime \prime }u^{\prime }-uu^{(4)}-\frac{1}{2}%
u^{^{\prime \prime }2}\right) \\
&  &  \\
a_{10} & = & {\frac{7}{256}}u^{5}+{\frac{35}{64}}{\theta }^{2}\left( \frac{1%
}{2}u^{2}u^{^{\prime }2}-u^{3}u^{\prime \prime }\right) +{\frac{7}{4}}{%
\theta }^{4}\left( \frac{3}{4}u^{(4)}u^{2}+\frac{7}{4}u^{^{\prime
\prime }2}u-\frac{3}{4}u^{\prime }{}^{2}u^{\prime \prime
}-uu^{\prime }u^{\prime
\prime \prime }\right) \\
&  &  \\
& + & {\frac{1}{4}}{\theta }^{6}\left( u^{\prime }u^{(5)}+\frac{1}{2}%
u^{\prime \prime \prime }{}^{2}-uu^{6}\right) \\
&  &  \\
a_{12} & = & -{\frac{21}{1024}}u^{6}+{\frac{105}{64}}{\theta
}^{2}\left(
u^{4}u^{\prime \prime }-\frac{1}{2}u^{3}u^{\prime }{}^{2}\right) \\
&  &  \\
& + & {\frac{1}{16}}{\theta }^{4}\left( 147uu^{\prime \prime
}u^{\prime
}{}^{2}+\frac{189}{2}u^{2}u^{\prime }u^{\prime \prime \prime }-\frac{1029}{4}%
u^{2}u^{^{\prime \prime }2}-63u^{3}u^{(4)}-\frac{105}{8}u^{\prime
}{}^{4}\right) \\
&  &  \\
& + & {\frac{1}{4}}{\theta }^{6}\left( 16u^{^{\prime \prime
}3}+9u^{2}u^{(6)}-27u^{\prime }u^{\prime \prime }u^{\prime \prime \prime }-%
\frac{45}{2}u^{\prime }{}^{2}u^{(4)}-\frac{69}{4}u^{^{\prime
\prime \prime }2}u+\frac{153}{2}uu^{\prime \prime
}u^{(4)}-\frac{27}{2}uu^{\prime
}u^{(5)}\right) \\
&  &  \\
& + & {\frac{1}{4}}{\theta }^{8}\left( u^{\prime
}u^{(7)}+u^{\prime \prime
\prime }u^{(5)}-u^{\prime \prime }u^{(6)}-uu^{(8)}-\frac{1}{2}{u^{(4)}}%
^{2}\right) \\
&  & \vdots
\end{array}
\end{equation}
with
\begin{equation}
a_{2k+1}=\sum_{s=0}^{2k-1}{\theta ^{s}}c_{2k-1}^{s}b_{2k+1-s}^{(s)}=0,%
\hspace{1cm}k=0,1,2,3,...
\end{equation}
Now having derived the explicit expression of ${}{\cal
L}^{\frac{1}{2}}$, we are now in position to write the explicit
forms of the set of $sl_{n}$-Moyal KdV hierarchy. These equations
defined as
\begin{equation}
\frac{\partial {}{\cal L}}{\partial t_{k}}=\{({}{\cal L}^{\frac{k}{2}%
})_{+},{}{\cal L}\}_{\theta },
\end{equation}
\newline
are computed in \cite{ref 23} up to the first three flows
$t_{1},t_{3},t_{5}$. We work out these equations by adding other
flows namely $t_{7}$ and $t_{9}$. We find
\begin{equation}
\begin{array}{lcl}
{u}_{t_{1}} & = & {u^{\prime }} \\
&  &  \\
{u}_{t_{3}} & = & \frac{3}{2}uu^{\prime }+{\theta ^{2}}u^{\prime
\prime
\prime } \\
&  &  \\
{u}_{t_{5}} & = & \frac{15}{8}u^{2}u^{\prime }+5{\theta
^{2}}(u^{\prime
}u^{\prime \prime }+\frac{1}{2}uu^{\prime \prime \prime })+{\theta }%
^{4}u^{(5)} \\
&  &  \\
{u}_{t_{7}} & = & {\frac{35}{16}}u^{3}u^{\prime }+\frac{35}{8}{\theta }^{2}({%
4}uu^{\prime }u^{\prime \prime }+u^{\prime 3}+u^{2}u^{\prime
\prime \prime })+\frac{7}{2}(uu^{(5)}+3u^{\prime
}u^{(4)}+{5}u^{\prime \prime }u^{\prime
\prime \prime }){\theta }^{4}+{\theta }^{6}u^{(7)} \\
&  &  \\
{u}_{t_{9}} & = & 18{\theta }^{6}u^{\prime }u^{(6)}+{\frac{651}{8}}{\theta }%
^{4}u^{\prime }(u^{\prime \prime })^{2}+{\frac{315}{128}}u^{4}u^{\prime }+{%
\frac{483}{8}}{\theta }^{4}u^{\prime }{}^{2}u^{\prime \prime \prime }+{\frac{%
315}{16}}{\theta }^{2}uu^{\prime }{}^{3}+{\frac{189}{4}}{\ \theta }%
^{4}uu^{(4)}u^{\prime } \\
&  &  \\
& + & {\frac{315}{8}}{\theta }^{2}u^{2}u^{\prime }u^{\prime \prime }+{\frac{%
315}{4}}{\theta }^{4}uu^{\prime }u^{\prime \prime \prime }+63{\theta }%
^{6}u^{\prime \prime \prime }u^{(4)}+\frac{105}{16}\theta
^{2}u^{3}u^{\prime
\prime \prime }+42{\theta }^{6}u^{(5)}u^{\prime \prime } \\
&  &  \\
& + & {\frac{63}{8}}{\theta }^{4}u^{2}u^{(5)}+{\theta }^{8}u^{(9)}+\frac{9}{2%
}{\theta }^{6}uu^{(7)}
\end{array}
\end{equation}
\newline
\newline
Some important remarks are in order:\newline
\newline
{\bf 1}. The flow parameters $t_{2k+1}$ has the following
conformal
dimension $[\partial _{t_{2k+1}}]=-[t_{2k+1}]=2k+1$ for $k=0,1,2,...,$.%
\newline
\newline
{\bf 2}. A remarkable property of the $sl_{2}$-Moyal KdV hierarchy
is about the degree of non linearity of the $\theta $-evolution
equations (60). We present in the following table the behavior of
the higher non-linear terms with respect to the first leading
flows $t_{1},...,t_{9}$ and give the behavior of the general flow
parameter $t_{2k+1}$.\newline
\newline
\begin{equation}
\begin{tabular}{ccc}
&  &  \\
&  &  \\
Flows & \hspace{1cm} The higher n.l. terms & \hspace{1cm} Degree
of n
linearity \\
$t_{1}$ & $u^{0}u^{\prime }=u^{\prime }$ & $0$ \\
&  &  \\
$t_{3}$ & $\frac{3}{2}uu^{\prime }$ & \hspace{2cm} $1$
\hspace{0cm}
(quadratic) \\
&  &  \\
$t_{5}$ & $\frac{15}{2^{3}}u^{2}u^{\prime }$ & \hspace{1.5cm} $2$
\hspace{0cm}
(cubic) \\
&  &  \\
$t_{7}$ & $\frac{35}{2^{4}}u^{3}u^{\prime }$ & $3$ \\
&  &  \\
$t_{9}$ & $\frac{315}{2^{7}}u^{4}u^{\prime }$ & $4$ \\
&  &  \\
... & ... & ... \\
&  &  \\
$t_{2k+1}$ & \hspace{0cm} $\eta {(2k+1)(2k-1)}u^{k}u^{\prime }$ & \hspace{1cm%
} $(k)$,
\end{tabular}
\end{equation}
where $\eta $ is an arbitrary constant.\newline
\newline
This result shows among others that the $\theta $-evolution
equations (60) exhibit at most a nonlinearity of degree $(k)$
associated to a term proportional to $(2k+1)(2k-1)u^{k}u^{\prime
}$. The particular case $k=0$ corresponds to linear wave
equation.\newline
\newline
{\bf 3}. The contribution of non-commutativity to the Moyal KdV
hierarchy shows a correspondence between the flows $t_{2k+1}$ and
the non-commutativity parameters $\theta ^{2(k-s)},0\leq s\leq k$.
Particularly, the higher term $\theta ^{2(k)}$ is coupled to the
$k-th$ prime derivative of $u_{2}$ namely $u^{(k)}$ while the
higher non linear term $\eta (2k+1)(2k-1)u^{k}u^{\prime }$ is a
$\theta $-independent object as its shown in (60).\newline
\newline
{\bf 4}. In analogy with the classical case, once the non linear
terms in the $\theta $-evolution equations are ignored, there will
be no solitons in the KdV-hierarchy as the latter's are intimately
related to non linearity.

\subsection{$sl_3$-Boussinesq Hierarchy}

The same analysis used in deriving the $sl_{2}$-KdV hierarchy is
actually extended to build the $sl_{3}$-Boussinesq Moyal
hierarchy. The latter is associated to the momentum Lax operator
${\cal L}{}_{3}=p^{3}+u_{2}\star p+u_{3}$ whose $3-th$ root reads
as
\begin{equation}
\begin{array}{lcl}
{\cal L}^{\frac{1}{3}} & = & \Sigma _{i=-1}b_{i+1}\star p^{-i} \\
&  &  \\
& = & {\Sigma }_{i=-1}a_{i+1}p^{-i}
\end{array}
\end{equation}
in such way that ${\cal L}{}_{3}={\cal L}{}^{\frac{1}{3}}\star {\cal L}{}^{%
\frac{1}{3}}\star {\cal L}{}^{\frac{1}{3}}$. Explicit computations
lead to
\begin{equation}
\begin{array}{lcl}
b_{0} & = & 1 \\
&  &  \\
b_{1} & = & 0 \\
&  &  \\
b_{2} & = & \frac{1}{3}u_{2} \\
&  &  \\
b_{3} & = & \frac{1}{3}u_{3}-\frac{2}{3}\theta u_{2}^{^{\prime }} \\
&  &  \\
b_{4} & = & -\frac{1}{9}u_{2}^{2}-\frac{2}{3}\theta u_{3}^{^{\prime }}+\frac{%
8}{9}\theta ^{2}u_{2}^{^{\prime \prime }} \\
&  &  \\
b_{5} & = & -\frac{2}{9}u_{2}u_{3}+\frac{8}{9}\theta u_{2}u_{2}^{\prime }+%
\frac{8}{9}\theta ^{2}u_{3}^{\prime \prime }-\frac{8}{9}\theta
^{3}u_{3}^{\prime \prime \prime } \\
&  &  \\
b_{6} & = & \frac{1}{9}\{\frac{5}{9}u_{2}^{3}-u_{3}^{2}+2\theta
(4u_{2}u_{3}^{\prime }+5u_{2}^{\prime }u_{3})-20\theta
^{2}(u_{2}u_{2}^{\prime \prime }+(u_{2}^{\prime })^{2})-8\theta
^{3}u_{3}^{\prime \prime \prime }+\frac{16}{3}\theta
^{4}u_{2}^{\prime
\prime \prime \prime }\} \\
&  &  \\
b_{7} & = & \frac{1}{9}\{\frac{5}{3}u_{2}^{2}u_{3}+10\theta
(u_{3}u_{3}^{\prime }-u_{2}^{2}u_{2}^{\prime })-\frac{20}{3}\theta
^{2}(5u_{2}^{\prime \prime }u_{3}+7u_{2}^{\prime }u_{3}^{\prime
}+u_{2}u_{3}^{\prime \prime \prime }) \\
&  &  \\
&  & -40\theta ^{3}(3u_{2}^{\prime }u_{2}^{\prime \prime
}+u_{2}u_{2}^{\prime \prime \prime })+\frac{16}{3}\theta
^{4}u_{3}^{\prime
\prime \prime \prime }) \\
&  &  \\
b_{8} & = & {\frac{5}{27}}(u_{{2}}u_{{3}}^{2}-{\frac{2}{9}}u_{{2}}^{4})-%
\frac{10}{9}\theta (u_{2}^{2}u_{3}^{\prime
}-\frac{7}{3}u_{2}^{\prime }u_{2}u_{3})+\frac{20}{81}\theta
^{2}(12u_{3}^{2\prime
}+31u_{2}u_{2}^{2\prime } \\
&  &  \\
&  & +17u_{2}^{2}u_{2}^{\prime \prime }-15u_{3}^{\prime \prime }u_{3})+\frac{%
40}{27}\theta ^{3}(10u_{3}^{\prime \prime }u_{2}^{\prime
}+13u_{2}^{\prime \prime }u_{3}^{\prime }+7u_{3}u_{2}^{\prime
\prime \prime }+3u_{3}^{\prime
\prime }u_{2}) \\
&  &  \\
&  & +\frac{80}{81}\theta ^{4}(8u_{2}^{4}u_{2}+23u_{2}^{2\prime
^{2}}+32u_{2}^{\prime }u_{2}^{\prime \prime \prime
})+{\frac{64}{81}}{\theta
}^{6}u_{{2}}^{(6)} \\
&  &
\end{array}
\end{equation}
Similarly, one can easily determine the coefficients $a_{i+1}$
which are also expressed as functions of $b_{i+1}$ and their
derivatives. This result is summarized in the expression of ${\cal
L}{}^{\frac{1}{3}}$ namely
\begin{equation}
\begin{array}{lcl}
{}{\cal L}^{\frac{1}{3}}=p & + & \frac{1}{3}u_{2}p^{-1} \\
&  &  \\
& + & \frac{1}{3}\{u_{3}-\theta u_{2}^{\prime }\}p^{-2} \\
&  &  \\
& - & \frac{1}{9}\{u_{2}^{2}+\theta ^{2}u_{2}^{\prime \prime }\}p^{-3} \\
&  &  \\
& + & \frac{1}{9}\{-2u_{2}u_{3}+2\theta u_{2}^{\prime
}u_{2}-\theta ^{2}u_{3}^{\prime \prime }+\theta ^{3}u_{2}^{\prime
\prime \prime }\}p^{-4}
\\
&  &  \\
& + & \frac{1}{9}\{\frac{1}{3}{\theta }^{4}u_{{2}}^{(4)}+2\theta u_{{2}%
}^{\prime }u_{{3}}-u_{{3}}^{2}+{\frac{5}{9}}u_{{2}}^{3}\}.{p}^{-5} \\
&  &  \\
& + & \frac{1}{27}\{{5}u_{{2}}^{2}u_{{3}}-{5}\theta u_{{2}}^{2}u_{{2}%
}^{\prime }+{10}{\theta }^{2}(u_{{2}}^{\prime }u_{{3}}^{\prime }-u_{{2}%
}^{\prime \prime }u_{{3}})+{\theta }^{4}u_{{3}}^{(4)}-{\theta }^{5}u_{{2}%
}^{(5)}\}.{p}^{-6} \\
&  &  \\
& + & \frac{1}{27}{\big \{ }\frac{5}{9}u_{2}(9u_{3}^{2}-2u_{2}^{3})-{10}%
\theta u_{{2}}^{\prime }u_{{2}}u_{{3}}+\frac{5}{3}\theta
^{2}(6u_{3}^{^{\prime }2}-6u_{3}^{\prime \prime
}u_{3}+5u_{2}^{2}u_{2}^{\prime \prime }-2u_{2}u_{2}^{2\prime }) \\
&  &  \\
& - & 10\theta ^{3}(-u_{3}^{\prime \prime }u_{2}^{\prime
}-u_{3}u_{2}^{\prime \prime \prime }+2u_{2}^{\prime \prime }u_{3}^{\prime })-%
\frac{10}{3}\theta ^{4}(u_{2}^{(4)}u_{2}+4u_{2}^{\prime
}u_{2}^{\prime
\prime \prime }-5u_{2}^{\prime \prime ^{2}})-{\frac{1}{3}}{\theta }^{6}u_{{2}%
}^{(6)}{\big \}}{p}^{-7} \\
&  &  \\
& + & ...
\end{array}
\end{equation}
Furthermore, using the Moyal $sl_{3}$-Lax evolution equations
\begin{equation}
\frac{\partial {}{\cal L}}{\partial t_{k}}=\{({}{\cal L}^{\frac{k}{3}%
})_{+},{}{\cal L}\}_{\theta },
\end{equation}
that we compute explicitly for $k=1,2,4$ we obtain
\begin{equation}
\begin{array}{lcl}
\frac{\partial {}{\cal L}}{\partial t_{1}} & = & u_{2}^{\prime
}p+u_{3}^{\prime }-{\theta }u_{2}^{\prime \prime } \\
&  &  \\
\frac{\partial {}{\cal L}}{\partial t_{2}} & = & 2\{u_{3}^{\prime
}-\theta
u_{2}^{\prime \prime }\}p-\frac{2}{3}\{u_{2}u_{2}^{\prime }+{\theta }%
^{2}u_{2}^{\prime \prime \prime }\} \\
&  &  \\
\frac{\partial {}{\cal L}}{\partial t_{4}} & = & \frac{4}{3}%
\{(u_{2}u_{3})^{\prime }-\theta (u_{2}^{\prime \prime
}u_{2}+u_{2}^{^{\prime }2})+2{\theta }^{2}u_{3}^{\prime \prime
\prime }-2{\theta }^{3}u_{2}^{(4)}\}p
\\
&  &  \\
&  & +\frac{4}{3}\{u_{3}u_{3}^{\prime
}-\frac{1}{3}u_{2}^{2}u_{2}^{\prime }-\theta (u_{2}^{\prime
}u_{3}^{\prime }+u_{2}^{\prime \prime }u_{3})-\theta
^{2}(u_{2}^{\prime }u_{2}^{\prime \prime }+u_{2}u_{2}^{\prime
\prime \prime
})-\frac{2}{3}\theta ^{4}u_{2}^{(5)}\} \\
&  &
\end{array}
\end{equation}
Identifying both sides of the previous equations, one obtain the
following first leading evolution equations
\begin{equation}
\begin{array}{lcl}
\frac{\partial }{\partial t_{1}}{u_{2}} & = & {u_{2}^{\prime }} \\
&  &  \\
\frac{\partial }{\partial t_{1}}{u_{3}} & = & {u_{3}^{\prime }} \\
&  &  \\
&  &  \\
\frac{\partial }{\partial t_{2}}{u_{2}} & = & 2{u_{3}^{\prime
}}-2\theta
u_{2}^{\prime \prime } \\
&  &  \\
\frac{\partial }{\partial t_{2}}{u_{3}} & = &
-\frac{2}{3}u_{2}u_{2}^{\prime }-\frac{8}{3}\theta
^{2}u_{2}^{\prime \prime \prime }+2\theta u_{3}^{\prime
\prime } \\
&  &  \\
&  &  \\
\frac{\partial }{\partial t_{4}}{u_{2}} & = & \frac{4}{3}\{(u_{2}u_{3})^{%
\prime }-\theta (u_{2}^{\prime \prime }u_{2}+u_{2}^{^{\prime }2})+2{\theta }%
^{2}u_{3}^{\prime \prime \prime }-2{\theta }^{3}u_{2}^{(4)}\} \\
&  &  \\
\frac{\partial }{\partial t_{4}}{(u_{3}-\theta u_{2}^{\prime })} & = & \frac{%
4}{3}\{u_{3}u_{3}^{\prime }-\frac{1}{3}u_{2}^{2}u_{2}^{\prime
}-\theta (u_{2}^{\prime }u_{3}^{\prime }+u_{2}^{\prime \prime
}u_{3})-\theta ^{2}(u_{2}^{\prime }u_{2}^{\prime \prime
}+u_{2}u_{2}^{\prime \prime \prime })-\frac{2}{3}\theta
^{4}u_{2}^{(5)}.
\end{array}
\end{equation}
These equations define what we call the Moyal $sl_{3}$ Boussinesq
hierarchy. The first two equations are simply linear $\theta
$-independent wave
equations fixing the dimension of the first flow parameter $t_{1}$ to be $%
[t_{1}]=-1$. \newline
\newline
The non trivial flow of this hierarchy starts really from the
second couple of equations associated to $t_{2}$. We will discuss
in the next section, how its important to deal with the basis of
primary conformal fields $v_{k}$ instead of the old basis $u_{k}$.
Anticipating this result, one can write
the previous couple of equations in term of the spin $3$ primary field $%
v_{3}=u_{3}-\theta u_{2}^{\prime }$ as follows
\begin{equation}
\begin{array}{lcl}
\frac{\partial }{\partial t_{2}}{u_{2}} & = & 2{v_{3}^{\prime }} \\
&  &  \\
\frac{\partial }{\partial t_{2}}{v_{3}} & = & -\frac{2}{3}%
\{u_{2}u_{2}^{\prime }+\theta ^{2}u_{2}^{\prime \prime \prime }\}
\end{array}
\end{equation}
This couple of equations define the $\theta $-extended Boussinessq
equation. Its second-order form is obtained by differentiating the
first equation in (68) with respect to $t_{2}$ and then using the
second equation. We find
\begin{equation}
\frac{\partial ^{2}}{\partial t_{2}^{2}}u_{2}=-\frac{4}{3}%
(u_{2}u_{2}^{\prime }+{\theta }^{2}u_{2}^{(3)})^{\prime },
\end{equation}
Equivalently one may write
\begin{equation}
\left( \matrix{u_2\cr v_3 \cr}\right)
_{_{t_{2}}}=-\frac{2}{3}\left( \matrix {-3v_{2}^{\prime }\cr
{u_{2}u_{2}^{\prime }+\theta ^{2}u_{2}^{\prime \prime \prime
}}\cr}\right).
\end{equation}
Similarly the third couple of equations (67) can be equivalently
written as
\begin{equation}
\begin{array}{lcl}
\frac{\partial }{\partial t_{4}}{u_{2}} & = &
\frac{4}{3}(u_{2}v_{3}+2\theta
^{2}v_{3}^{\prime \prime })^{\prime } \\
&  &  \\
\frac{\partial }{\partial t_{4}}{v_{3}} & = & \frac{4}{3}\{v_{3}v_{3}^{%
\prime }-\theta ^{2}u_{2}u_{2}^{\prime \prime \prime }-\frac{1}{3}%
(u_{2}^{2}u_{2}^{\prime }+2\theta ^{4}u_{2}^{(5)})\}
\end{array}
\end{equation}
Recall that the classical Boussinesq equation is associated to the $sl_{3}$%
-Lax differential operator
\begin{equation}
{\cal L}{}_{3}=\partial ^{3}+2u\partial +v_{3}
\end{equation}
with $v_{3}=u_{3}-\frac{1}{2}u_{2}^{\prime }$ defining the
spin-$3$ primary field. This equation which takes the following
form
\begin{equation}
u_{tt}=-(auu^{\prime }+bu^{(3)})^{\prime },
\end{equation}
where $a,b$ are arbitrary constants, arises in several physical
applications. Initially, it was derived to describe propagation of
long waves in shallow water \cite{ref 24}. This equation plays
also a central role in $2d$ conformal field theories via its
Gelfand-Dickey second Hamiltonian structure associated to the
Zamolodchikov $w_{3}$ non linear algebra \cite{ref 17, ref
18}.\newline
\newline
To close this section note that other flows equations associated to $%
(sl_{2}) $-KdV and $(sl_{3})$-Boussinesq hierarchies can be also
derived once some lengthly and hard computations are performed.
One can also generalize the obtained results by considering other
$sl_{n}$ integrable hierarchies with $n>3$.
\section{The NC Lax generating technics in the Moyal momentum framework}
Using the convention notations and the analysis presented
previously and developed in \cite{ref 20, ref 21} and based also
on the results established in \cite{ref 25, ref 26, ref
27}\footnote{I am grateful to K. Toda for bringing to my attention
ref. 27}, we present in this section some results related to the
Lax representation of noncommutative integrable hierarchies. We
perform also consistent algebraic computations, based on the
Moyal-momentum analysis, to derive explicit Lax pair operators of
some integrable systems in
the noncommutative framework.\\\\
We underline that the present formulation is based on the (pseudo)
momentum operators $p^n$ ad $p^{-n}$ instead of the (pseudo)
operators $\partial^n$ and $\partial^{-n}$ used in several works.
We note also that the obtained results are shown to be compatible
with the ones already established in literature \cite{ref 26}.\\\\
Note also that the notion of integrability of the concerned
nonlinear differential equations is defined in the sense that
these equations may be linearizable.\\\\
To start, let's recall that the $sl_n$-Moyal KdV hierarchy is
defined as
\begin{equation}
\frac{\partial {\mathcal L}}{\partial t_{k}}=\{({\mathcal
L}^{\frac{k}{2}})_{+} ,{\mathcal L}\}_{\theta}.
\end{equation}
Explicit computations related to these hierarchies are presented
in \cite{ref 20, ref 21}. Working these hierarchies, we was able
to derive, among others, for the $sl_2$ case up to the flow $t_9$,
the following KdV-hierarchy equations
\begin{equation}
\begin{array}{lcl}
{u}_{t_{1}}&=&{u'},\\\\
{u}_{t_{3}}&=&\frac{3}{2}uu'+{\theta^2}u''',\\\\
{u}_{t_{5}}&=&\frac{15}{8}u^{2}u'+5{\theta^2}(u'u''+\frac{1}{2}uu''')+{\theta}^{4}u^{(5)},\\\\
{u}_{t_{7}}&=&{\frac{35}{16}} u^{3}u^{\prime }+\frac{35}{8}{\theta }%
^{2} ( {4} uu^{\prime }u^{\prime \prime }+ u^{\prime 3}+
u^{2}u^{\prime \prime \prime } ) +\frac{7}{2}( uu^{(5)}+3
u^{\prime }u^{(4)}+{5} u^{\prime \prime }u^{\prime \prime \prime }
) {\theta }^{4}+{\theta }^{6}u^{(7)}, \\\\
...
\end{array}
\end{equation}
Actually this construction which works well for the $sl_2$-KdV
hierarchy is generalizable to higher order KdV hierarchies, namely
the $sl_n$-KdV hierarchies.\\\\
The basic idea of the Lax formulation consists first in
considering a noncommutative integrable system which possesses the
Lax representation such that the following noncommutative Moyal
bracket
\begin{equation}
\{L, T+\partial_t\}_{\theta}=0,
\end{equation}
is equivalent to the noncommutative differential equation that we
consider from the
beginning and that is nonlinear in general with $\partial_{t}\equiv \frac{\partial}{\partial t}$.\\
Equation (76) and the associated pair of operators $(L,T)$ are
called the Lax differential equation and the Lax pair,
respectively. The differential operator $L$ defines the integrable
system which we should fix from the beginning.\\\\
Note that the way with which ones to writes the Lax equation as in
(76) is equivalent to that in (74) namely
\begin{equation}
\{L, T+\partial_t\}_{\theta}\equiv\{L, ({\mathcal
L}^{\frac{k}{2}})_{+}+\partial_{t_k}\}_{\theta}=0,
\end{equation}
where the operator $T$ is the analogue of $({\mathcal
L}^{\frac{k}{2}})_{+}$ describing then an operator of conformal
spin $k$.\\\\
This equation, written in terms of the function $u(x,t)$, is in
general a non linear differential equation belonging to the ring
${\widehat{\Sigma} _{k+2}^{(0,0)}}$. In the present case of
$sl_2$-KdV systems we have $k=3$.\\\\
As it's shown in  \cite{ref 26}, the meaning of Lax
representations in noncommutative spaces would be vague. However,
they actually have close connections with the bi-complex method
\cite{ref 28} leading to infinite number of conserved quantities,
and the (anti)-self-dual Yang-Mills equation which is integrable
in the context of twistor descriptions and ADHM constructions
\cite{ref 29, ref 30}.
\\\\ Now, let us apply the noncommutative Lax-pair generating technique.
Usually, it's a method to find a corresponding $T$-operator for a
given $L$-operator. Finding the operator $T$ satisfying (76) is
not an easy job in the general case. For this reason, one have to
make
some constraints on the operator $T$ namely:\\\\
{\bf Ansatz for the operator $T$}:
\begin{equation}
T=p^n\star L^{m}+ T',
\end{equation}
where $p^n$ are momentum operators acting on arbitrary function
$f(x,p)$ as shown in \emph{section 2}. Note by the way that the
notation $T'$ have nothing to do with the prime derivative. With
the previous ansatz, the problem reduces to that for the
$T'$-operator which is determined by hand so that the Lax equation
should be a differential equation
belongings to the ring ${\widehat{\Sigma}^{(0,0)}}$.\\
The best way to understand what happens for the general case, is
to focus on the following examples:\\
{\bf Example 1: The $sl_2$-noncommutative KdV system.}\\\\
The $L$-operator for the noncommutative KdV equation is given, in
the momentum space configuration, by
\begin{equation}
L=p^2+u(x,t),
\end{equation}
with
\begin{equation}
L\in {\widehat{\Sigma} _{2}^{(0,2)}}/{\widehat{\Sigma}
_{2}^{(1,1)}},
\end{equation}
where ${\widehat{\Sigma} _{2}^{(1,1)}}$ is the one dimensional
subspace generated by objects of type $\xi_{1}(x,t)\star p$ and
$\xi_{1}(x,t)$ is an arbitrary function of conformal spin $1$.\\\\
Reduced to $n=1=m$, for the noncommutative $sl_2$ KdV system, the
ansatz (25) can be written as follows\footnote{One can also
introduce the following definition: $T\equiv (p\star L)_{s}+ T'$,
with the convention $(p\star L)_{s}\equiv \frac{(p\star L+L\star
p)}{2}$ describing the symmetrized part of the operator $p\star
L$}
\begin{equation}
T=p\star L+ T'.
\end{equation}
The operator $T$ in this case ($k=3$), is shown to behaves as
$({\mathcal L}^{\frac{3}{2}})_{+}$ with $\partial_{t_3}\equiv
\frac{\partial}{\partial t_3}$.\\\\
Simply algebraic computations give
\begin{equation}
\{L, T'\}_{\theta}=u'\textbf{p}^2-2\theta
u''\textbf{p}+\frac{\dot{u}}{2\theta}+(uu'+\theta^2u'''),
\end{equation}
Next, our goal is to be able to extract the Lax differential
equation, namely, the noncommutative KdV equation. Before that, we
have to make a projection of the operator $\{L, T'\}_{\theta}$ on
the ring ${\widehat{\Sigma} _{3+2}^{(0,0)}}$. This projection is
equivalent to cancel the effect of the terms of momentum in (82),
namely the term $u'\textbf{p}^2$ and $2\theta
u''\textbf{p}$. To do that, we have to consider the following property: \\\\
{\bf Ansatz for $T'$}:\\
\begin{equation}
T'=X\star p+ Y,
\end{equation}
where $X$ and $Y$ are arbitrary functions on $u$ and its
derivatives.
\\\\
Next, performing straightforward computations, with $T'=Xp-\theta
X'+Y$ lead to
\begin{equation}
\{L,T'\}_{\theta}=2X'\textbf{p}^{\textbf{2}}+(\{u,X\}_{\theta}-2\theta
 X''+2Y')\textbf{p}+ (-Xu'-\theta \{u,X'\}+\{u,Y\} )
\end{equation}
Identifying (82) and (84), leads to the following constrains
equations
\begin{equation}
X=\frac{1}{2}u_{2}+a,
\end{equation}
\begin{equation}
Y=-\frac{1}{2}\theta u_{2}'+b,
\end{equation}
 with the following nonlinear differential equation
\begin{equation}
-\frac{\dot{u}}{2\theta}=\frac{3}{2}uu'+\theta^{2}u'''.
\end{equation}
where the constants $a$ and $b$ are to be omitted for a matter of
simplicity. The last equation is noting but the $sl_2$ KdV
equation. This noncommutative equation contains also a non linear
term $\frac{3}{2}uu'$.\\\\We have to underline that the $sl_2$
noncommutative KdV equation obtained through this Lax method
belongs to the same class of the KdV equation derived in \cite{ref
20, ref 21} namely
\begin{equation}
\dot{u}=\frac{3}{2}uu'+\theta^{2}u'''.
\end{equation}
In fact, performing the following scaling transformation
$\partial_{t_3}\rightarrow -2\theta\partial_{t_3}$ we recover
exactly (88). The term $\frac{1}{2\theta}$ appearing in (87) as
been the coefficient of the evolution part $\dot{u_{2}}$ of the NC
KdV equation can be simply shifted to one due to consistency with
respect to the classical limit $\theta_{l} \sim \frac{1}{2}$.
\\\\ To summarize, the momentum Lax pair operators, associated to the
noncommutative $sl_2$-KdV system, are explicitly given by
\begin{equation}
L_{KdV}=p^{2}+u_{2}(x,t),
\end{equation}
and
\begin{equation}
T_{KdV}=\textbf{p}^3+\frac{3}{2}\textbf{p}\star
u_{2}(x,t)-\frac{3}{2}\theta u_{2}'(x,t);
\end{equation}
with $T'=\frac{1}{2}\textbf{p}\star u_{2}(x,t)-\frac{3}{2}\theta u'_{2}(x,t)$\\\\
Note that, the same results can obtained by using the
Gelfand-Dickey (GD) formulation based on formal (pseud)
differential operators $\partial ^{\pm n}$ instead of the Moyal
momentum
ones.\\\\
This first example shows, among others, the consistency of the
Moyal momentum formulation in describing integrable systems and
the associated Lax pair generating technics in the same way as the
successful GD formulation \cite{ref 16}.\\\\
{\bf Example 2: The Noncommutative Burgers Equation $\grave a $ la
Moyal}\\
Let us apply the same noncommutative Lax technics, presented
previously, to derive the noncommutative version of the Burgers
equation. Actually, our interest in this equation comes from the
several important properties that are exhibited in the commutative
case. Before going into applying the noncommutative Lax technics,
let's first recall some few known properties of the standard
Burgers
equation.\\\\
{\bf P1}: The Burgers equation is defined on the $(1+1)$-
dimensional space time. In the standard pseudo-differential
operator formalism, this equation is associated to the following
L-operator
\begin{equation}
L_{Burgers}=\partial_x+u_{1}(x,t)
\end{equation}
where the function $u_1$ is of conformal spin one. Using our
convention notations, we can set $L\in {\widehat
\Sigma}^{(0,1)}_{1}$.\\
{\bf P2}: With respect to the previous L-operator, the non linear
differential equation of the Burgers equation is given by
\begin{equation}
\dot{u}_{1}+\alpha u_{1}u'_{1}+\beta u''_{1}=0,
\end{equation}
where $\dot{u}=\frac{\partial u}{\partial t}$ and
$u'=\frac{\partial u}{\partial x}$. The dimensions of the
underlying objects are given by $[t]=-2=-[\partial_t]$, $[x]=-1$
and $[u]=1$.\\\\
{\bf P3}: On the commutative space-time, the Burgers equation can
be derived from the Navier-Stokes equation and describes real
phenomena, such as the turbulence and shock waves. In this sense,
the Burgers equation draws much attention amongst many integrable
equations.\\\\
{\bf P4}: It can be linearized by the Cole-Hopf transformation
\cite{ref 31}. The linearized equation is the diffusion equation
and can be solved by Fourier transformation for given boundary
conditions.\\\\
{\bf P5}: The Burgers equation is completely integrable \cite{ref 32}.\\\\
Now, we are ready to look for the noncommutative version of the
Burgers equation. For that, we consider the $L$-operator of this
equation in the Moyal momentum language, namely
\begin{equation}
L=p+u_{1}(x,t).
\end{equation}
dealing, as noticed before, to the space ${\widehat
\Sigma}^{(0,1)}_{1}$. This is a local differential operator of the
generalized $n$-KdV hierarchy's family $(n=1)$, obtained by a
truncation of a noncommutative pseudo momentum operator of KP
hierarchy type
\begin{equation}
L=p+u_{1}(x,t)+u_{2}(x,t)\star p^{-1}+u_{3}(x,t)\star p^{-2}+...,
\end{equation}
of the space ${\widehat \Sigma}^{(-\infty,1)}_{1}$. The local
truncation is simply given by
\begin{equation}
{\widehat \Sigma}^{(-\infty,1)}_{1}\rightarrow {\widehat
\Sigma}^{(0,1)}_{1}\equiv [{\widehat
\Sigma}^{(-\infty,1)}_{1}]_{+}\equiv {\widehat
\Sigma}^{(-\infty,1)}_{1}/{\widehat \Sigma}^{(-\infty,-1)}_{1},
\end{equation}
or equivalently
\begin{equation}
L_{1}(u_i)=p+\Sigma _{i=0}^{\infty}u_{i}\star p^{1-i} \rightarrow
p+u_{1}\equiv[L_{1}(u_i)]_{+},
\end{equation}
where the symbol $(X)_{+}$ defines the local part (only positive
powers of $p$) of a given pseudo operator $X$.
\\\\
The noncommutative Burgers equation is said to have the Lax
representation if there exists a suitable pair of operators
$(L,T)$ so that the Lax equation
\begin{equation}
\{p+u_1, T+\partial_t\}_{\theta}=0,
\end{equation}
reproduces the noncommutative version of the Burgers non linear
differential equation. Following the same steps developed
previously for the $sl_2$ noncommutative KdV systems, we consider
the following ansatz for the operator $T$:
\begin{equation}
T=p\star L+ T',
\end{equation}
or
\begin{equation}
T=p^2+ u_{1}p+ \theta u'_{1}+ T'.
\end{equation}
Then, performing straightforward computations, the noncommutative
Burgers Lax equation (97) reduces to
\begin{equation}
\{p+u,T'\}_{\theta}=u'\textbf{p}+(uu'-\theta
u''+\frac{\dot{u}}{2\theta})
\end{equation}
Next, one have also the go through a constraint equation for the
operator $T'$, namely the
\\\\
{\bf Ansatz for $T'$}:
\begin{equation}
T'=A\star p+ B,
\end{equation}
where $A$ and $B$ are arbitrary functions for the moment.With this
new ansatz for $T'$, we have
\begin{equation}
\{p+u,T'\}_{\theta}=(A'+\{u,A\}_{\theta})\textbf{p}+(-Au'-\theta
A''-\theta \{u,A'\}_{\theta}+B'+\{u,B\}_{\theta})
\end{equation}
\\Identifying (100) and (102) leads to the following constraints
equations
\begin{equation}
u'=A'+\{u,A\}_{\theta})
\end{equation}
and
\begin{equation}
(u+A)u'+\frac{\dot{u}}{2\theta}=B'+\{u,B\}_{\theta}+\theta
\{A',u\}+\theta(u''-A'')
\end{equation}
A natural solution of the first constraint equation (103) is
$A=u$. This leads to a reduction of (104) to
\begin{equation}
2uu'+\frac{\dot{u}}{2\theta}=B'+\{u,B\}
\end{equation}
Actually this is the noncommutative Burgers equation, which is
also the projection of the Lax equation (97) to the ring of
vanishing degrees in momenta namely the space ${\widehat{\Sigma}
_{1}^{(0,0)}}$.\\\\
Since $\{u,\partial_t\}_{\theta}=-\frac{\dot u}{2\theta}$, a non
trivial solution of the parameter $B$ in equation (105) consists
in setting $B\equiv u^2-\frac{\partial}{\partial t}$. But, since
this nontrivial solution of $B$ masks the noncommutative Burgers
equation, it's a non desirable thing.\\\\
Remarking also that $[B]=2$, we use this dimensional arguments and
set
\begin{equation}
B=\xi u'+\eta u^2
\end{equation}
with $\xi$ and $\eta$ are arbitrary coefficient numbers. Putting
this expression into (105) gives the final expression of the
noncommutative Burgers equation namely
\begin{equation}
\frac{\dot{u}}{2\theta}+2(1-\eta)uu'-\xi u''=0
\end{equation}
whose Lax pair in the noncommutative Moyal momentum formalism are
explicitly given by
\begin{equation}
L_{Burgers}=p+u_{1}(x,t)
\end{equation}
and
\begin{equation}
T_{Burgers}=p^2+2u_{1}(x,t)p+\eta u_{1}^2(x,t)+\xi u_{1}'(x,t)
\end{equation}
\section {\textbf{Noncommutative Burgers-KdV mapping}}
This section will be devoted to another significant aspect of the
noncommutative integrable models.  The principal focus, for the
moment, is on the models discussed previously namely the
noncommutative KdV and Burgers systems.  In \emph{section 5} we
discussed the integrability of these two nonlinear systems and we
noted that they are indeed integrable and this property is due to
the existence of definite Lax pair operators $(L, T)$ for each of
the two models. Such existence implies the linearization of the
models automatically.\\\\
A crucial question which arises now is to know if it there is a
possibility to establish a mapping between the two Systems.  The
idea to connect the two models is originated from the fact that
integrability for the KdV system both in commutative and
noncommutative spaces is something natural due to the possibility
to connect with $2d$ conformal symmetry. We think that the strong
backgrounds of conformal symmetry can help to shed more lights
about integrability of the noncommutative Burgers systems if one
know how to establish such a connection.\\\\
On the other hand, it is clear that these models are different due
to the fact that for the noncommutative KdV system the Lax
operator as well as the associated field $u_{2}(x,t)$ are of
conformal weights $2$, whereas for the Burgers system, $L$ and $
u_1$ are of weight $1$.
\\\\Our goal is to study the possibility of transition between the
two spaces ${\widehat \Sigma}_{2}^{(0,2)}/{\widehat
\Sigma}_{2}^{(1,1)}$ and ${\widehat \Sigma}_{1}^{(0,1)}$
corresponding respectively to the two models.  This transition,
once it exists, should leads to extract more informations on these
noncommutative models and also on their integrability.\\\\ To
start, let's consider the following property
\\\\
{\bf Proposition 1:}\\ Lets consider the Burgers momentum operator
$L_{Burgers}(u_1)=p+u_{1}\in {\widehat \Sigma}_{1}^{(0,1)}$. For
any given $sl_2$ noncommutative KdV operator
$L_{KdV}(u_2)=p^2+u_{2}(x,t)$ belongings to the space ${\widehat
\Sigma}_{2}^{(0,2)}/{\widehat \Sigma}_{2}^{(1,1)}$, one can define
the following mapping
\begin{equation}
{\widehat \Sigma}_{1}^{(0,1)} \hookrightarrow {\widehat
\Sigma}_{2}^{(0,2)}/{\widehat \Sigma}_{2}^{(1,1)},
\end{equation}
in such away that
\begin{equation}
L_{Burgers}(u_1)\rightarrow L_{KdV}(u_2)\equiv
L_{Burgers}(u_1)\bigotimes L_{Burgers}(-u_1).
\end{equation}\\
We know that the space ${\widehat \Sigma}_{2}^{(0,2)}$ of momentum
operators of conformal spin $s=2$ is different from the one of
momentum operators of conformal spin $s=1$ namely ${\widehat
\Sigma}_{1}^{(0,1)}$. What we are assuming in this proposition is
a strong constraint leading to connect the two spaces. This
constraint is also equivalent to set
\begin{equation}
{\widehat \Sigma}_{2}^{(0,2)}/{\widehat \Sigma}_{2}^{(1,1)}\equiv
{\widehat \Sigma}_{1}^{(0,1)}\bigotimes {\widehat
\Sigma}_{1}^{(0,1)}
\end{equation}
Next we are interested in discovering the crucial key behind the
previous proposition. For this reason, we underline that this
mapping is easy to highlight through the noncommutative analogue
of the well known Miura transformation
\begin{equation}
L_{KdV}=p^2+u_{2}=(p^1+u_{1})\star (p^1-u_{1})
\end{equation}
giving rise to
\begin{equation}
u_{2}=-u_{1}^{2}-2\theta u_{1}^{'}.
\end{equation}
This is an important property since one have the possibility to
realize the KdV $sl_2$ noncommutative field $u_2$ in term of the
Burgers field $u_1$, its derivative $u'_{1}$ and of the
$\theta$-parameter. This realizations shows among other an
underlying nonlinear behavior in the KdV noncommutative field
$u_{2}$ given by the quadratic term $u^{2}_{1}$.
\\\\However, the \emph{proposition 1} can have a complete and
consistent significance only if one manages to establish a
connection between the noncommutative differential equations
associated to the two systems.  Arriving at this stage, note that
besides the principal difference due to conformal spin, we stress
that the two nonlinear evolutions equations of KdV
\begin{equation}
-\frac{1}{2\theta}\frac{\partial u_2}{\partial
t_{3}}=\frac{3}{2}uu'+\theta^{2}u'''.
\end{equation}
and of Burgers
\begin{equation}
\frac{1}{2\theta}\frac{\partial u_1}{\partial
t_{2}}+2(1-\eta)uu'-\xi u''=0
\end{equation} are distinct by a
remarkable fact that is the KdV flow $t_{KdV} \equiv t_3$ and the
Burgers one $t_{Burgers}\equiv t_2$ have different conformal
weights: $[t_{KdV}]=-3$ whereas $[t_{Burgers}]=-2$.
\\\\In order to be consistent with the objective of the \emph{proposition 1}, based on the idea of the possible link
between the two noncommutative integrable systems, presently we
are constrained to circumvent the effect of proper aspects
specific to both the equations and consider the following second
property:
\\\\
{\bf Proposition 2:}\\\\
By virtue of the Burgers-KdV mapping and dimensional arguments,
the associated flow are related through the following ansatz
\begin{equation}
\partial_{t_2}\hookrightarrow
\partial_{t_3}\equiv\partial_{t_2}.\partial_{x}+\alpha
\partial_{x}^{3}
\end{equation}
for an arbitrary parameter $\alpha$.
\\\\With respect to the assumption (117), relating the two
evolution derivatives $\partial_{t_2}$ and $\partial_{t_3}$
belongings to Burgers and KdV hierarchies respectively, one should
expect some strong constraint on the Burgers noncommutative
differential equation (116). Such constraint is important since
one needs to fix the arbitrary coefficients $\xi$ and $\eta$ which
are still
arbitrary.\\
We have to identify the following three differential equations
\begin{equation}
\begin{array}{lcl}
\partial_{t_3}u_2&=&\frac{3}{2}u_{2}u'_{2}+\theta^{2}u'''_{3},\\\\
&=&-2u_{1}\partial_{t_3}u_{1}-2\theta \partial_{t_3}u'_{1},\\\\
&=&\partial_{t_2}u'_{2}+\alpha u'''_{2}.
\end{array}
\end{equation}
{\bf Some explicit results}:\\\\
Setting for a matter of simplicity the Burgers equation as
$\partial _{t_{2}}u_{1}=Au_{1}u'_{1}+Bu^{''}_{1}$ with
$A=4\theta(\eta-1)$ and $B=2\theta \xi$, and performing explicit
computation, rising from the identification of the previous system of equations (118), we find the following results:\\\\
\begin{equation}
\begin{array}{lcl}
\partial_{t_3}u_2&=&3u^{3}_{1}u_{1}'+6\theta u^{'2}_{1}u_{1}+3\theta
u^{''}_{1}u_{1}^{2}-2\theta^{2}u^{'''}_{1}u_{1}-2\theta^{3}u^{''''}_{1}\\\\
&=&-2Au^{'2}_{1}u_{1}-6\theta
Au^{''}_{1}u'_{1}-2Au^{2}_{1}u''_{1}-2\theta
u^{''''}_{1}(B+\alpha)-2u_{1}u^{'''}_{1}(\alpha+A\theta+B )\\\\
&=&-4Au^{'2}_{1}u_{1}-2u^{''}_{1}u'_{1}(B+3\alpha+3A\theta)-2u^{'''}_{1}u_{1}(B+\alpha+A\theta)-2Au''_{1}u^{2}_{1}
-2\theta u^{''''}_{1}(B+\alpha)
\end{array}
\end{equation}
These expressions, once are simplified, lead to the following
constraint equation
\begin{equation}
Au_{1}u'_{1}+(B+3\alpha)u''_{1}=0.
\end{equation}
Putting this constraint equation into the noncommutative Burgers
equation (116) give the following equation
\begin{equation}
\partial_{t_2}u_{1}=-3\alpha u''_{1},
\end{equation}which is a linear differential equation.
This is an impressing result deserving special interest since one
have the possibility to convert a nonlinear differential
equation to a linear one.\\\\
We will try now to explain the introduced Burgers-KdV mapping and
the induced linearizability property in connection with our guess
of a hidden $2d$-conformal symmetry. First of all note that the
conformal symmetry in the framework of noncommutative KdV
hierarchy is related in general to the $sl_n$-symmetry as it's
explicitly shown in \emph{section {3}}, see also \cite{ref 20} for
more details. Referring to these algebraic backgrounds, we can
make contact with Burgers-KdV mapping. In fact, we have to remark
that the Burgers $u_1$-current issued from the Miura like equation
(113) and satisfying (114) can be identified, by virtue of
(45-46), with the Liouville Lorentz scalar field $\phi$ as follows
\begin{equation}
u_{1}(x,t)\equiv \phi^{'}
\end{equation}
with $\phi^{'}\equiv \partial \phi\equiv\frac{\partial}{\partial
x}\phi(x,t)$ while the noncommutative KdV potential $u_{2}(x,t)$
satisfying (114) can be then identified with the conformal current
$T$ given by (47). Using all these equations one can actually
interpret the constraint equation (120), induced from the
Burgers-KdV mapping, as been the equation of conservation of the
noncommutative KdV potential $u_{2}$. In fact, making an analogy
with $2d$ conformal field theory construction and using (114) the
noncommutative KdV current read in terms of the $u_1$-Burgers
current as $T\equiv u_{2}(x,t)=-u_{1}^{2}-2\theta u_{1}^{'}$ which
coincides also with (47) once we introduce the $\theta$-Liouville
field $\phi$. Requiring the conservation property for this
current, namely
\begin{equation}
u_{1}u'_{1}+\theta u_{1}^{''}=0
\end{equation}
reproduce exactly the constraint equation (120) induced from the
Burgers-KdV mapping with the following fixation of the parameters
$\xi$ and $\eta$
\begin{equation}
\begin{array}{lcl}
\xi&=& \frac{1}{4\theta}+1\\ \\
\eta&=&\frac{\theta-3\alpha}{2\theta}
\end{array}
\end{equation}
On the other hand, the constraint equation (120), which we
consider now as been a conservation law, is also equivalent to set
\begin{equation}
\phi=-\theta log{\phi{''}},
\end{equation}
giving rise then to the following $\theta$-Liouville like
differential equation
\begin{equation}
{\phi}{''}=Kexp(-\frac{1}{\theta}\phi)
\end{equation}
Based on the analogy with (48) the constant can be simply chosen
equal to $K=\frac{2}{\theta}$
\newpage
\section{Concluding Remarks}
1. The results obtained for the Moyal Momentum
algebra\footnote{The appellation of Moyal momentum algebra
introduced for the first time by Das and Popowicz, see \cite{ref
33, ref 34}}are applied to study some properties of $sl_{2}$-KdV
and $sl_{3}$-Boussinesq integrable hierarchies. Our contributions
to this study consist in extending the results found in literature
by increasing the order of computations a fact which leads us to
discover more important properties as its explicitly shown in
\cite{ref 20, ref 21}.\\\\
2. We have presented a systematic study of the Moyal momentum
algebra that we denote in our convention notation as
${\widehat{\Sigma }}_{\theta }$. This is the huge space of
momentum Lax operators of arbitrary conformal spin $m$,$m\in Z$
and arbitrary higher and lowest degrees $(r,s)$ reading as
\begin{equation}
{\tilde{{\cal L}}}_{m}^{(r,s)}(u)=\sum_{i=r}^{s}p^{i}\star u_{m-i}
\end{equation}
3. We studied the algebraic properties of ${\widehat{\Sigma
}}_{\theta }$ and its underlying sub-algebras ${\widehat{\Sigma
}}_{m}^{(r,s)}$ and show that among all these spaces only the
subspace ${\widehat{\Sigma }}_{0}^{(-\infty ,1)}$ which defines a
Lie algebra structure with respect to the Moyal bracket.\\\\
4. The particular sub-algebra $sl_{n}-{\widehat{\Sigma
}}_{n}^{(0,n)}$ built out of the $sl_{n}$ momentum Lax operators ${\tilde{{\cal L}}}%
_{n}^{(0,n)}(u)=\sum_{i=0}^{n}p^{i}\star u_{n-i}$, with $u_{0}=1$ and $%
u_{1}=0$, is applied to field theory building. Indeed, using the
properties of this sub-algebra we were able to construct the
$\theta $-Liouville conformal model
\begin{equation}
\partial {\bar{\partial}}\phi =\frac{2}{\theta }e^{-\frac{1}{\theta }\phi }
\end{equation}
and its $sl_{3}$-Toda extension.
\begin{equation}
\begin{array}{lcl}
\partial {\bar{\partial}}\phi _{1} & = & Ae^{-\frac{1}{2\theta }(\phi _{1}+%
\frac{1}{2}\phi _{2})} \\
&  &  \\
\partial {\bar{\partial}}\phi _{2} & = & Be^{-\frac{1}{2\theta }(\phi
_{1}+2\phi _{2})}
\end{array}
\end{equation}
5. We show also that the central charge, a la Feigin-Fuchs,
associated to the spin-2 conformal current of the $\theta
$-Liouville model is given by
\begin{equation}
c_{\theta }=(1+24\theta ^{2})
\end{equation}
6. We derived also the noncommutative $sl_{2}$-KdV and
$sl_3$-Boussinesq hierarchies and write their associated
$\theta$-flows. The NC KdV equation is given by
\begin{equation}
\dot{u}=\frac{3}{2}uu'+\theta^{2}u''',
\end{equation}
while the NC Boussinesq equation given by the couple of equations
\begin{equation}
\left( \matrix{u_2\cr v_3 \cr}\right)
_{_{t_{2}}}=-\frac{2}{3}\left( \matrix {-3v_{2}^{\prime }\cr
{u_{2}u_{2}^{\prime }+\theta ^{2}u_{2}^{\prime \prime \prime
}}\cr}\right)
\end{equation}
\\
7. Besides the above established results in the Moyal momentum
framework, we tried also to understand much more the meaning of
integrability of noncommutative nonlinear systems. The principal
focus was on the NC KdV and NC Burgers systems. For this reason, a
first contribution was to derive these two equations using the
above systematic algebraic formulation in the context of noncommutative Lax pair building (\emph{Section
5}).\\\\
8. Concerning the noncommutative derived KdV system, this is an
integrable model due to the existence of a noncommutative Lax pair
operators $(L, T)$. This existence is an important indication of
integrability, but we guess that the realistic source of
integrability of this model is the underlying conformal symmetry,
shown to play a similar role as in the commutative case. The
derived NC KdV equation $-\frac{1}{2\theta}{\partial_{t_{3}}
u_2}=\frac{3}{2}uu'+\theta^{2}u'''$, through the NC Lax pair
building, is equivalent to the one derived in \emph{Section 4}
\cite{ref 20, ref 21}
${\partial_{t_{3}}u_2}=\frac{3}{2}uu'+\theta^{2}u'''$ once the
scaling transformation $\partial_{t_3}\rightarrow
-2\theta\partial_{t_3}$ is considered.\\\\
9. Concerning the noncommutative Burgers system (107) that we
consider in the second example, it's also an integrable equation
whose Lax pair operators are explicitly derived (108-109). Note
for instance that the Burgers Lax operator
$L_{Burgers}=p+u_{1}(x,t)$ is a momentum operator belongings to
the space ${\widehat\Sigma}_{1}^{(0,1)}$.\\\\
10. As a first checking of integrability for the noncommutative
Burgers system, we proceeded to an explicit derivation of the Lax
pair operators $(L,T)$ giving the following differential equation
$2uu'+\frac{\dot{u}}{2\theta}=B'+[u,B]$. The idea is to solve this
equation in terms of the coefficient parameter $B$ such that it
can reduces to the non linear Burgers equation belonging to the
space ${\widehat\Sigma}_{3}^{(0,0)}$. Solving this equation give
explicitly the requested Lax operator.\\\\
11. We should also underline that the importance of this study
comes also from the fact that the results obtained in the
framework of Moyal momentum are similar to those coming by
using the Gelfand-Dickey pseudo operators approach \cite{ref 26, ref
27}.\\\\
12. Concerning the possibility to establish a correspondence
between the NC KdV and NC Burgers systems. Actually, we succeeded
to build a mapping leading to transit from the NC Burgers system
to the NC KdV system. The main line of this mapping deals with the
following ansatz $\partial_{t_2}\hookrightarrow
\partial_{t_3}\equiv\partial_{t_2}\partial_{x}+\alpha
\partial_{x}^{3}$ for an arbitrary parameter $\alpha$. This ansatz is important for several reasons, we give
here bellow some of them:\\\\
{\bf a.)} It implies the linearization of the Burgers equation.\\\\
{\bf b.)} It helps to fix the arbitrary NC Burgers coefficients
$\xi$ and $\eta$ through a strong linearizability constraint
$Au_{1}u'_{1}+(B+3\alpha)u''_{1}=0$.\\\\
{\bf c.)} From the conformal field theory point of view, this
constraint equation is nothing but the analogue of the
conservation law of the conformal current.\\\\ {\bf d.)} The idea
behind this mapping, as it's shown in \emph{Section 6}, is that
for the noncommutative KdV model, the problem of integrability
does not arise in the same way as it's for the noncommutative
Burgers equation. The first one is mapped to conformal field
theory through the Liouville model. This is in fact a strong
indication of integrability in favor of
noncommutative KdV equation which could help more to understand the noncommutative Burgers system.\\
We believe that the considered mapping might help to bring new
insights towards understanding the integrability of noncommutative
$2d$-systems.\\\\
{\bf Acknowledgements} \\\\
The author would like to thank the Abdus Salam International
Center for Theoretical Physics (ICTP) for hospitality and
acknowledge the considerable help of the High Energy Section and
its head S. Randjbar-Daemi. This work was done at ICTP, within the
framework of the Associateship Scheme of the Abdus Salam
International Center for Theoretical Physics, Trieste, Italy. I am
thanking the Associateship office and more particularly its staff
for the quality of service and for the permanent availability. I
present special acknowledgements to OEA-ICTP and to its head
George Thompson for valuable scientific helps in the context of
Network. I also wish to thank the University Ibn Tofail, the
Faculty of Sciences at Kenitra and the protas III programm D12/25
CNR, Morocco. Best acknowledgements are also presented to H. Saidi
for valuable scientific discussions.

\end{document}